# A Flexible Semi-Automatic Approach for Glioblastoma multiforme Segmentation

J. Egger, M. H. A. Bauer, D. Kuhnt, C. Kappus, B. Carl, B. Freisleben and C. Nimsky

*Abstract*—Gliomas are the most common primary brain tumors, evolving from the cerebral supportive cells. For clinical follow-up, the evaluation of the preoperative tumor volume is essential. Volumetric assessment of tumor volume with manual segmentation of its outlines is a time-consuming process that can be overcome with the help of segmentation methods. In this paper, a flexible semi-automatic approach for grade IV glioma segmentation is presented. The approach uses a novel segmentation scheme for spherical objects that creates a directed 3D graph. Thereafter, the minimal cost closed set on the graph is computed via a polynomial time s-t cut, creating an optimal segmentation of the tumor. The user can improve the results by specifying an arbitrary number of additional seed points to support the algorithm with grey value information and geometrical constraints. The presented method is tested on 12 magnetic resonance imaging datasets. The ground truth of the tumor boundaries are manually extracted by neurosurgeons. The segmented gliomas are compared with a one click method, and the semi-automatic approach yields an average Dice Similarity Coefficient (DSC) of 77.72% and 83.91%, respectively.

## I. INTRODUCTION

Gliomas are primary brain tumors, evolving from the cerebral supportive cells. Their incidence in Europe is 7-11/100000 inhabitants, thus infrequent as compared to other human malignancies. The type is determined by the cells they arise of, most frequently astrocytomas (astrocytes), oligodendrogliomas (oligodendrocytes) or ependymomas (ependymal) cells. Furthermore, there are mixed forms containing different cell types, such as oligoastrocytomas. With over 60%, astrocytic tumors are the most common tumors. The grading system for astrocytomas according to the World Health Organization (WHO) subdivides grades I-IV, whereas grade I tumors tend to be least aggressive [1]. According to its histopathological appearance, the highly malignant grade IV tumor is given the name glioblastoma multiforme (GBM). It is the most frequent glioma with approximately 50%, followed by astrocytomas WHO I-III with 25%. Oligodendrogliomas and Ependymomas are less frequent with 5-18% and 2-9%, respectively.

The glioblastoma multiforme is one of the highest malignant human neoplasms, although they lack metastases. The interdisciplinary therapeutical management today contains maximum safe resection, percutaneus radiation and most frequently, chemotherapy. Still, despite new radiation strategies and the development of oral alcylating substances (e.g. Temozolomide), the survival rate is still only approximately 15 months [2]. In former years, the surgical role was controversial, since a direct association between the amount of tumor resection and patient survival was questioned. Recently, several studies have been able to prove maximum surgical resection as a positive predictor for patient survival [3]. Today, this microsurgical procedure is optimized with the technical development of neuronavigation containing functional data sets such as diffusion tensor imaging (DTI), functional MRI (fMRI), magnetoencephalography (MEG), magnetic resonance spectroscopy (MRS), or positron-emission-computed-tomography (PET).

For clinical follow-up, the evaluation of the tumor volume in the course of disease is essential. The volumetric assessment of a tumor applying manual segmentation is a time-consuming process. In this paper, a semi-automatic graph-based segmentation approach for World Health Organization grade IV gliomas is presented. The approach uses a user-defined seed point inside the tumor to create a spherical, directed graph. Then, the minimal cost closed set on the graph is computed via a polynomial time s-t cut, providing an optimal segmentation of the tumor. The center of the polyhedron is defined by the user and located inside the glioma. However, the user can improve the results by specifying an arbitrary number of additional seed points to support the algorithm. To evaluate the novel approach, WHO IV grade gliomas are manually, semi-automatically and automatically (one click) segmented in magnetic resonance imaging (MRI) datasets. The results of the (semi-) automatic segmentations are compared with manual segmentations of three neurological surgeons who are experts in their fields and have several years of experience in the resection of gliomas. The Dice Similarity Coefficient (DSC) is calculated to evaluate the outcomes.

The paper is organized as follows. Section 2 reviews related work. Section 3 presents the details of the proposed approach. In Section 4, experimental results are discussed. Section 5 concludes the paper and outlines areas for future work.

## II. RELATED WORK

For glioma segmentation based on MRI, several algorithms have already been proposed. Angelini et al. [4] give a broad overview of some deterministic and statistical approaches. The majority of them are region-based approaches; more recent ones are based on deformable models and include edge-

J.E., M.H.A.B., D.K., C.K., B.C., Ch.N. are with the Department of Neurosurgery, University of Marburg, Marburg, Germany e-mail: egger@med.uni-marburg.de

J.E., M.H.A.B., B.F. are with the Department of Mathematics and Computer Science, University of Marburg, Marburg, Germany

information.

Gibbs et al. [5] have presented a combination of region growing and morphological edge detection for segmentation of enhancing tumors in T1 weighted MRI data. Starting with a manually provided first sample of tumor signal and surrounding tissue, an initial segmentation using pixel thresholding, morphological opening and closing and fitting to an edge map is performed. Gibbs et al. have evaluated their procedure with one phantom dataset and ten clinical datasets. However, the mean segmentation time for a tumor was about ten minutes, and they did not exactly classify the tumors they used for their evaluation.

Letteboer et al. [6] have proposed an interactive method for segmentation of full-enhancing, ring-enhancing and non-enhancing tumors. They evaluated their approach with twenty clinical cases. Based on a manual tracing of an initial slice, morphological filter operations are applied to the MRI volume to divide the data in homogenous regions.

Depending on intensity-based pixel probabilities for tumoral tissue, a deformable model has been presented by Droske et al. [7]. They use a level set formulation, in order to split the MRI data into regions of similar image properties for tumor segmentation. The method was then performed on image data of twelve patients.

A knowledge-based automated segmentation on multispectral data to partition glioblastomas has been introduced by Clark et al. [8]. After a training phase with fuzzy C-means classification, clustering analysis and a brain mask computation, initial tumor segmentation from vectorial histogram thresholding is postprocessed to eliminate non-tumor pixels. The introduced system has been trained on three volume data sets and has been tested on thirteen unseen volume data sets.

Prastawa et al. [9] have proposed a segmentation based on outlier detection in T2 weighted MRI data. Therefore, in order to detect abnormal tumor regions, the image data is registered on a normal brain atlas. Then, tumor and edema are isolated by statistical clustering of the differing voxels and a deformable model. However, they have applied the method only to three real data sets. For each case, the required time for automatic segmentation was about 90 minutes.

Sieg et al. [10] have introduced an approach for segmenting contrast-enhanced, intracranial tumors and anatomical structures of registered multisprectral MRI data. Multilayer feedforward neural networks with backpropagation are trained and a pixel-oriented classification is applied. The approach has been tested on twenty-two data sets, but the authors did not provide any computational time.

### III. METHODS

The proposed approach uses a novel segmentation scheme for spherical objects that creates a directed 3D graph by (1) sending rays through the surface points of a polyhedron and (2) sampling the graph's nodes along every ray. Thereafter, the minimal cost closed set on the graph is computed via a polynomial time s-t cut [11], creating an optimal segmentation of the tumor. The center of the polyhedron is defined by the user and located inside the glioma.

To set up the graph, the method samples along rays that are sent through the surface points of a polyhedron with the user-defined seed point as the center (Figure 1). The sampled points are the nodes $n \in V$ of the graph $G(V,E)$ and $e \in E$ is a set of edges. There are edges between the nodes (Figure 2) and edges that connect the nodes to a source $s$ and a sink $t$ to allow the computation of an $s$-$t$ cut.

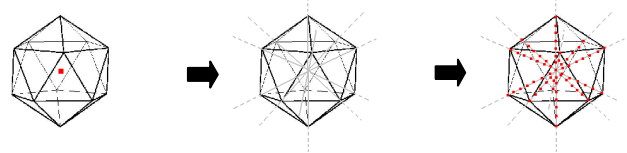

Fig. 1. Principle of sampling the graph's nodes (red points in the rightmost image).

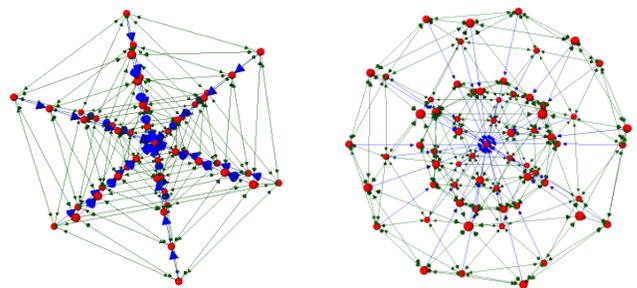

Fig. 2. Principle of graph construction. 5 (left) and 3 (right) sampled points (red) along each of the 12 (left) and 32 (right) rays that provide the nodes for the graph.

The idea of setting up the graph with a polyhedron goes back to a catheter simulation algorithm where several polyhedra were used to align the catheter inside the vessel [12]. In the novel segmentation scheme, this idea is combined with a graph-based method that has been introduced for the semi-automatic segmentation of the aorta [13], [14], [15] and diffusion tensor imaging (DTI) fiber bundle segmentation [16]. However, the user can enhance the results by specifying an arbitrary number of additional seed points to support the algorithm with grey value information and geometrical constraints.

The average grey value that is needed for the calculation of the costs and the graph's weights (for details see [15] and [16]) is essential for the segmentation result. Based on the assumption that the user-defined seed points are inside or at least on the border of the tumor, the average grey value can be estimated automatically. Therefore, we integrate over small cubes of dimension d centered on the user-defined seed points $(s_x, s_y, s_z)$, with $s$ as the number of seed points:

$$\frac{\sum_{i=1}^{s} \int_{-d/2}^{d/2} \int_{-d/2}^{d/2} \int_{-d/2}^{d/2} T(s_x+x, s_y+y, s_z+z)dxdydz}{s} \quad | \; s > 0 \quad (1)$$

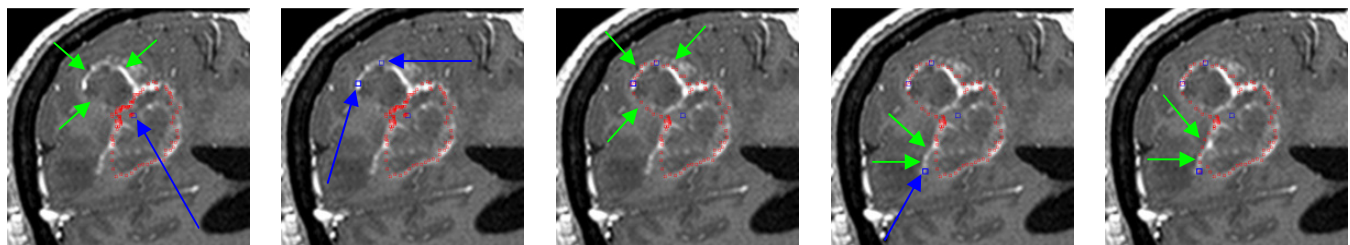

Fig. 3. Principle of the additional seed points that support the one click algorithm.

Additionally, the extra seed points ensure that the resulting surface contains these user-defined restrictions. To make this behavior possible, the rays that contain an additional user-defined seed point are set "fixed". This is realized by binding all "following" nodes of a considered ray with maximum weight to the source and all "previous" nodes – including the additional seed point – with maximum weight to the sink. This graph construction forces the mincut to follow the user's input for the optimal s-t cut.

## IV. RESULTS

To evaluate the approach, three neurological surgeons with several years of experience in the resection of tumors performed manual slice-by-slice segmentation of 12 WHO grade IV gliomas. The tumor outlines for the segmentation were displayed by the contrast-enhancing areas in T1 weighted MR images. Afterwards, the segmentation results were compared with the one-click-segmentation results of the proposed method via the Dice Similarity Coefficient (DSC) [17].

The Dice Similarity Coefficient is the relative volume overlap between A and R, where A and R are the binary masks from the automatic (A) and the reference (R) segmentation. $V(\bullet)$ is the volume (in $cm^3$) of voxels inside the binary mask, by means of counting the number of voxels, then multiplying with the voxel size. The average Dice Similarity Coefficient for all data sets was 77.72% – for the one click method without additional seed points – and 83.91% – for the semi-automatic approach supporting the algorithm by 15 to 75 extra seed points – (see Table I).

Figure 3 (upper left) shows the result of a segmented tumor border (red) with the one click method. The segmentation started from the user-defined seed point (blue arrow). However, a border area was not segmented by the algorithm (green arrows). The next two images (in the upper row of Figure 3) show two additional seed points by the user (blue arrows) and the result of the algorithm (green arrows). It is evident that the missed tumor border can be segmented with the two additional seed points. Equivalently, the algorithm was supported by an extra seed point in the lower left corner (see two rightmost images in Figure 3). In Figure 4, several segmentation results of the presented approach are shown as 3D models.

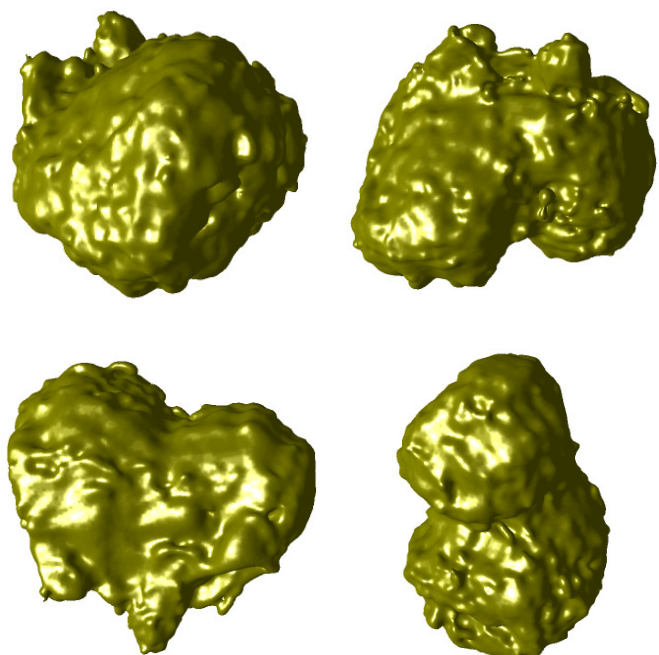

Fig. 4. Four 3D models of (semi-) automatically segmented tumors.

The presented methods were implemented in C++ within the medical image processing platform MeVisLab [18]. Using 2432 and 7292 polyhedra surface points, the overall segmentation (sending rays, graph construction and mincut computation) in our implementation took less than 30 seconds on an Intel Core i5-750 CPU, 4x2.66 GHz, 8 GB RAM, Windows XP Professional x64 Version, Version 2003, Service Pack 2.

TABLE I
SUMMARY OF RESULTS: MIN., MAX., MEAN AND STANDARD DEVIATION FOR 12 GLIOMAS.

|  | Volume of tumor ($cm^3$) | | | Number of voxels | | | $DSC_{oneClick}$ (%) | $DSC_{semi}$ (%) |
| --- | --- | --- | --- | --- | --- | --- | --- | --- |
|  | manual | one click | semi | manual | one click | semi | | |
| min | 2.38 | 0.99 | 0.99 | 2694 | 993 | 993 | 53.76 | 72.61 |
| max | 86.91 | 65.6 | 67.51 | 283011 | 252316 | 252316 | 91.96 | 91.96 |
| $\mu \pm \sigma$ | 21.02 ± 26.48 | 15.55 ± 20.18 | 20.31 ± 22.16 | 57278.33 | 45461.50 | 58196.67 | 77.72 ± 13.19 | 83.91 ± 6.91 |

## V. CONCLUSION

In this paper, a flexible semi-automatic approach for World Health Organization grade IV gliomas (Glioblastoma multiforme) has been introduced. The presented approach uses a new segmentation scheme for spherically shaped objects and creates a 3D graph in two stages: In the first stage, rays are sent through the surface points of a polyhedron, and in the second stage, the graph's nodes are sampled along every ray. Afterwards, the minimal cost closed set on the constructed graph is computed via a polynomial time s-t cut, resulting in an optimal segmentation of the tumor boundaries. The center of the polyhedron is user-defined and located inside the tumor. Additionally, the user can enhance the segmentation scheme by specifying an arbitrary number of seed points that support the algorithm with grey value information and geometrical constraints. These "extra" seed points given by the user ensure that the resulting surface – describing the tumor – contains these user-defined restrictions. The presented algorithm has been evaluated on 12 magnetic resonance imaging datasets with WHO grade IV gliomas. Three experts (neurosurgeons) with several years of experience in resection of gliomas extracted the tumor boundaries manually to obtain the ground truth of the data. The manually segmented results, a one click method – with no extra seed points besides the polyhedron center – and the semi-automatic approach – supporting the algorithm by 15 to 75 additional seed points have been compared with each other by calculating the average Dice Similarity Coefficient.

There are several areas of future work. For example, the presented segmentation scheme can be enhanced with statistical information about shape and texture of the desired object [19]. Furthermore, the method can be evaluated on MRI datasets with WHO grade I, II and III gliomas. Additionally, the approach can be used to segment other pathologies like cerebral aneurysms [20] and spherically shaped organs.

## VI. ACKNOWLEDGEMENTS

The authors would like to thank Fraunhofer MeVis in Bremen, Germany, for their collaboration and especially Horst K. Hahn for his support.